# The brain is a computer is a brain: neuroscience's internal debate and the social significance of the Computational Metaphor

**Authors:** Alexis T. Baria[1]* and Keith Cross[2]

**Affiliations:**
[1]*Society of Spoken Art, New York, NY, USA.*
[2]*Curriculum Studies Department, College of Education, University of Hawai`i at Manoa, Honolulu, HI, USA.*
*\*Correspondence to Alexis T. Baria, email: alexis.t.baria@gmail.com.*

## Abstract

The Computational Metaphor, comparing the brain to the computer and vice versa, is the most prominent metaphor in neuroscience and artificial intelligence (AI). Its appropriateness is highly debated in both fields, particularly with regards to whether it is useful for the advancement of science and technology. Considerably less attention, however, has been devoted to how the Computational Metaphor is used outside of the lab, and particularly how it may shape society's interactions with AI. As such, recently publicized concerns over AI's role in perpetuating racism, genderism, and ableism suggest that the term "artificial intelligence" is misplaced, and that a new lexicon is needed to describe these computational systems. Thus, there is an essential question about the Computational Metaphor that is rarely asked by neuroscientists: whom does it help and whom does it harm? This essay invites the neuroscience community to consider the social implications of the field's most controversial metaphor.

## Introduction

Science fiction would have us understand the potential dangers of AI in the form of machines that, with their apparent ability to "out-think" and/or "out-perform" humans, have "decided" to colonize or exterminate humans, (a historically human inclination). Science fact regarding AI is quite different, but with potentially similar outcomes -- humans who, to their own peril, have been convinced by other enterprising humans to place excessive faith in AI that cannot "out-think" or "out-perform" them except in a few respects that are yet poor substitutes for human's intellectual and other expressly human (at least for the time being) capacities. Last year, for example, OpenAI's latest language model, GPT-3, was tested as a viable healthcare chatbot, and promptly suggested that a fake patient should commit suicide because they "felt bad" (Daws, 2020). While in this instance GPT-3 did not perform as hoped, large language models in general are fluent enough to give a false impression of language understanding and mental modeling (Bender et al., 2021), thus epitomizing the concept of "artificial intelligence". Cases like the above, however, call into question whether the intelligence and brain-based terminology used to market AI technology poses risks for the parties upon or by whom the technology is used.

At the core of this terminology is perhaps the most debated metaphor in all of science, the Computational Metaphor (Biron, 1993; Cisek, 1999; Cobb, 2020; Karaliutė, 2009; MacCormac, 1984; Marcus, 2015;

Marshall, 1977; Richards, 2018; Smith, 1993; Vlasits, 2017; West & Travis, 1991). Drawing parallels between computer and brain functions, neuroscientists frequently conceptualize that THE BRAIN IS A COMPUTER. Indeed, for neuroscientists, the implication of the Computational Metaphor is quite clear: it is needed to advance the field because scientific metaphors, in general, are valuable tools for explaining complex subject matter and generating useful ideas (Taylor & Dewsbury, 2018). But the Computational Metaphor is bigger than neuroscience -- it is encountered in everyday life and implied in every mention of AI. Thus, as highly critical users of that metaphor, neuroscientists shape how it is communicated both in and outside of the lab.

The purpose of this essay is not to argue whether the brain is a computer or not. Rather, it is to point out that the Computational Metaphor (which is debated frequently and publicly by brain researchers) has tangible and important social ramifications which have received little attention from brain researchers. Instead, as debates about the metaphor's academic utility go on, artificial intelligence, whose label itself invokes the Computational Metaphor, has been shown to perpetuate social harms that are often at the expense of those who are under-represented in these debates, and in academia / the tech industry, in general (Allyn, 2020; Angwin et al., 2016; Benjamin, 2019; Birhane & Guest, 2020; Buolamwini & Gebru, 2018; Crawford et al., 2019; O'Neil, 2017; Ong, 2017; Stark, 2019). We believe unchecked use of the Computational Metaphor contributes to these harms by falsely attributing human-like capabilities to AI-labeled technologies, and aiding in a disregard for the complexity of social and human experiences. Neuroscientists and brain researchers should then consider a crucial question when centering their work around the Computational Metaphor: whom does it benefit, and whom does it harm?

## The Computational Metaphor

According to conceptual metaphor theory, metaphor is not just a matter of language, but a matter of thought (Lakoff & Johnson, 1980). The Computational Metaphor thus affects how people understand computers and brains, and of more recent importance, influences interactions with AI-labeled technology -- this is the central concern of our argument. We will break down the metaphor below, and show how the concepts embedded within it afford the human mind less complexity than is owed, and the computer more wisdom than is due. We will notate other key metaphors as well in all-caps, and will discuss the Computational Metaphor in two ways: THE BRAIN IS A COMPUTER, and THE COMPUTER IS A BRAIN.

The first form of the metaphor, THE BRAIN IS A COMPUTER, is frequently debated by neuroscientists regarding its utility to advance science; while the second form, THE COMPUTER IS A BRAIN, is often how technologists and tech advocates communicate AI to non-experts. Each form of the Computational Metaphor is the reverse direction of the other, and each is meant to serve very different purposes, as is common according to theories of metaphor directionality (Turner, 1998). Thus, a neuroscientist may use THE BRAIN IS A COMPUTER to explain how retinas transduce light into electrical signals by comparing it to the image sensor of a camera. Conversely, a computer engineer could use THE COMPUTER IS A BRAIN to explain random access memory in terms of human working memory. Both directions of the Computational Metaphor share the common purpose of explaining something abstract in terms of something concrete. And as with many metaphors, each direction is meant to offer specific and distinct understandings. Take for instance the common metaphor LOVE IS A JOURNEY -- the forward direction means something entirely different, in terms of experience and sensibility, from the reverse, A JOURNEY IS LOVE.

The Computational Metaphor, however, is not like common unidirectional metaphors. It is a special case of metaphor whose forward and reverse meanings can become entangled, and as such its bidirectional properties have been of interest for decades (Karaliutė, 2009; MacCormac, 1984; Stein, 1999; West & Travis, 1991). Why exactly the Computational Metaphor behaves this way is not entirely clear, but it is likely due to the long history of medical and scientific metaphors comparing humans to various engineered objects (Marshall, 1977; Smith, 1993), and the history of the term "computer", which originally signified a person who performed calculations. Moreover, neuroscience terminology has been used to describe modern computers for nearly 80 years: titles of popular computer publications in the 1940s include "Electrical Memory", "It Thinks With Electrons", "100-Ton Brain at MTI" (Barry, 1993), and "Giant Brains or Machines That Think" (Berkeley, 1949). These historical associations are seeds of thought, deeply buried in our language and grown over time. This poses a challenge for neuroscientists who model their work around THE BRAIN IS A COMPUTER, and wish to untangle and distinguish the roots of their thoughts from those supporting the notion that THE COMPUTER IS A BRAIN.

Prominent computer scientists have raised concerns about the entangled meanings that emerge from this bidirectionality, suggesting that treating machines like people can lead to treating people like machines. Joseph Weizenbaum, the developer of the chatbot Eliza, was one of the first prominent critics of computer anthropomorphization for this reason (Weizenbaum, 1976). And Edsger W. Dijkstra, who coined the phrase, "structured programming" also spoke on his concerns about the unintended reversal of the Computational Metaphor:

*"A more serious byproduct of the tendency to talk about machines in anthropomorphic terms is the companion phenomenon of talking about people in mechanistic terminology. The critical reading of articles about computer-assisted learning... leaves you no option: in the eyes of their authors, the educational process is simply reduced to a caricature, something like the building up of conditional reflexes. For those educationists, Pavlov's dog adequately captures the essence of Mankind —while I can assure you, from intimate observations, that it only captures a minute fraction of what is involved in being a dog,"* (Dijkstra, 1985).

The bidirectionality of the Computational Metaphor is also apparent in our everyday conversations. Consider the following figures of speech which entail THE BRAIN IS A COMPUTER: "I can't *process* all that information"; "Let me *crunch the numbers*"; "You can *ping* me later"; "He doesn't have the *bandwidth* for this". The reverse, THE COMPUTER IS A BRAIN, is also quite common: "My computer is *sleeping*"; "The upgraded model has tons of *memory*"; "The camera *sees* my face"; "My laptop won't *talk* to the projector". These examples are so unremarkable it is easy to feel that they are not metaphorical at all. Thus, it is important for scientists to keep in mind that scientific metaphors can alter everyday language and understandings of phenomena to the point that the sense of metaphor (and the distinguishing features of its components) disappears.

In fact, the distinction between metaphor and literality has been yet another concern among computer scientists, arguing that the perceived exactness of the Computational Metaphor not only limits scientific creativity, but also limits human attributes:

*"Metaphors can be most dangerous when one forgets they are metaphors; one can become beguiled by familiarity rather than by corroborating evidence into accepting a metaphor as literal ... leading us to assume that the attributes normally possessed by either referent are possessed in the same way by the other. If humans and computers both possess 'beliefs', then a person may be led by the metaphorical usage to assume that the properties of human 'belief' should be limited to dispositions to act, since they are so limited in computers,"* (MacCormac, 1984).

Yet the literal take on the Computational Metaphor seems quite popular as of recent (Marcus, 2015), and the mathematical reasoning is convincing (Richards, 2018). The general argument is that brains are not equal to laptops or smartphones; but rather that brains, laptops, and smartphones all fall under an abstract, mathematical definition of a computer. While this may be literally true, it doesn't align with how most people understand computers, and thus this brain/computer comparison still risks being widely mishandled. To put it another way, from a public outreach standpoint, it would be problematic for an institution to campaign the literal statement, *a gun is a hole-punch*er. This is accurate, and can be taken as literal, insomuch as we define hole-punchers as objects that put holes in things. One could even study the properties of the hole punching element to design a better bullet, or vice versa. But one should re-evaluate the cost of canvassing that statement if people begin perforating their paper documents with rifles.

This is not to say that comparing brains and computers with guns and hole-punchers is of equal utility or absurdity, but only that literally true statements can still be interpreted in unintended ways, and just because they might be true doesn't mean they don't require careful communication. When scientists deliver their messages to the public, they do it from a position of influence, and this poses an extra challenge which can be overlooked (or sometimes brushed off) in the desire to avoid over-simplifying for a non-expert audience. To downplay the importance of how the public understands the Computational Metaphor, however, is to forget that basic scientific research is largely funded by the public. As such, scientists are obligated to ensure that highly influential work is communicated in a language that is not easy to abuse. Moreover, it should be accessible to the audience that pays for that messaging, who may not have the resources to pick apart the intricate arguments that can mislead even those with a privileged education in the topic.

## The social influence of scientific metaphor

While those with educational access and deep knowledge about brains and computers might understand why the Computational Metaphor is or is not technically accurate, this understanding remains mostly invisible to everyone else. This allows the Computational Metaphor to be transformed in unintended ways, such that the human mind is afforded less complexity than is owed, and the computer is afforded more wisdom than is due. With such a powerful idea embedded in so much of our daily interactions with technology, the Computational Metaphor risks perpetuating harmful messages which can impact social policy and behavior (Hartung et al., 2020; Taylor & Dewsbury, 2018).

The label, "artificial intelligence", invokes the Computational Metaphor, and carries this risk. As such, it has some AI researchers concerned about how the general population interprets this label, and the downstream effects of that interpretation. Human-computer interaction scholar, Meredith Ringel Morris, had this to say about it in a recent podcast:

*"Frankly, I am bothered by the terminology 'artificial intelligence'. This goes back to my earlier point about setting expectations and how we communicate to the public at large about our work. When a layperson hears the term 'artificial intelligence' they think we're talking about what computer scientists would call 'artificial general intelligence'. They think of something that has a human-like intelligence, a semantic intelligence. Whereas current trends in machine learning are all about pattern recognition and statistics and no semantic understanding and knowledge… We should actually be more concerned about these pattern-based definitions of ML that lack semantics and therefore result in this kind of inadvertent bias and ethical issues... I wish we didn't have the term 'intelligence' in AI because what computer scientists mean by 'intelligence' and what regular people mean by 'intelligence' is just not the same thing"* (Doyle-Burke & Smith, 2021).

This divergent framing of "intelligence" between experts and non-experts matters. Framing even basic concepts, like weight, with different terms can lead to dramatically different perceptions (Slepian & Ambady, 2014). It is understandable then, that complicated social phenomena, like intelligence, are subject to the same influence. Studies have shown that single phrases have this effect on social issues: crime as a "virus" vs "beast" (Thibodeau & Boroditsky, 2011); the climate crisis as a "war" rather than a "race" (Flusberg et al., 2018); and genetically altered food as "engineered" rather than "modified" (Zahry & Besley, 2019); all have been shown to shift people's attitudes on important and touchy subjects. Because computers and AI-labeled technologies are such an integral part of daily life, how the Computational Metaphor is communicated and used is a social issue, and likewise may have similar effects on social behavior and policy.

One example is last year's botched rollout of COVID-19 vaccines at Stanford Medical Center. When front-line workers began protesting hospital administrators having first access to the vaccine, administrators blamed it on an "algorithm". Most people took this term to mean that the decisions were made by machine learning, when really it was just a relatively simple formula designed by a few people (Lum & Chowdhury, 2021). Nonetheless, healthcare workers gathered in the lobby of the medical center yelling "Fuck the *algorithm*" (emphasis ours), seemingly directing their anger at a mathematical procedure. This calls to question whether "algorithm" was chosen intentionally over other terms, and if so, why. As the studies above might suggest, is there something about AI-terminology that changes perceptions on whether decisions are good or bad? Would protesters have been chanting something more appropriate like "fuck the administrators", if a term that is less likely to evoke the concept of AI, like "formula" or "procedure", been used? More importantly, would workers have felt more agentive in those cases to contest the decisions being made on their behalf?

Coming back to the term, "artificial intelligence", AI researcher, Kate Crawford, posits that "AI is neither artificial nor intelligent" (Corbyn, 2021). We also see both these terms as central to widespread misconceptions about AI-labeled technology, as they seem to promote undue trust. In education, the ways in which intelligence is defined and measured is understood to vary with culture (Sternberg, 2015) and context (Gardner & Hatch, 1989; Pea, 1993). In principle, artificial intelligence should not be exempt from this rule, and yet there are specific ways in which it is. Even taken in parts, AI defies common sense. Artificial flavors are generally assumed to be fake flavors and substandard to authentic flavors, a chemical trick to add to the shelf life of a food product or fool the end user's taste buds. Somehow, the term, "artificial intelligence", turns the common sense regarding artifice on its head, by invoking the ideologies of the superiority of rational thought. Returning briefly to education, one such ideology ranks people as more or less intelligent according to their mastery of mathematics, and it is regularly assumed that people who are "good" at math are of superior intellect compared to those who are not "good" at math. So prominent is this understanding that people in the latter category learn to avoid math at all costs, to avoid bringing attention to what they have been conditioned to believe is an inferior intellectual ability. Add to this that computers seem to be really good at math, and we begin to understand the social power of the Computational Metaphor. Although we might consider being mercurial (i.e. calculating and without emotion) as an undesirable human personality trait, the Computational Metaphor rests on other well-ingrained ideologies in which a hierarchy of human value is tied to a particular notion of intelligence such that the quality of being emotional is considered inferior to being rational. This notion of intelligence extends to the justified subjugation of beings considered less rational to those considered (or propagandized as) more rational, whether animals to humans, women to men, or one race of humans to another. According to this logic, in its *fake-ness* as a human intelligence, AI paradoxically succeeds in being a more trustworthy form of intelligence, by being the epitome of rational thought.

The Computational Metaphor reifies this trust, and it can be invoked, as above for instance, to deter recourse and feign accountability in ways less brain-centric terminology might not. And while an important goal in technology is to build responsible products that are worthy of trust, it has been shown that trust in artificial intelligence can actually lead people to make choices which put themselves in danger (Robinette et al., 2016; Wagner et al., 2018). Use of the Computational Metaphor, then, might also be considered with respect to how it influences trust in AI-labeled tech, and in turn how that tech may be used to control the systems which are dependent on that trust. We discuss that more, below.

## The brain and AI as control systems

The brain (James, 1890) and AI are classically described as input-output boxes, a key similarity upon which the Computational Metaphor is constructed. In order for either to use information about their surroundings, input data has to come in, and it has to be of a certain form. Sound waves, for example, only activate perception in the brain within a limited range of frequencies. Likewise, AI algorithms only work with data that is just so -- but how that data relies on social groundwork to become a usable form is often overshadowed by marvels of model size and complexity. Usually, a series of human interventions first needs to be performed before input data can be used by an AI algorithm: collecting, cleaning, transforming, and labeling are all crucial steps in getting an algorithm to distinguish patterns in data. A common estimate in the data science community is that data preparation is at least 80% of the total workload in making a machine learning model. Prior to this process, however, there's social infrastructure and human labor required for data collection -- work that is performed by content-moderators, platform and freelance workers, warehouse workers, artists, social media users, and nearly anyone with a connected device. But despite their absolute necessity, human input, intent, and labor are rarely acknowledged as part of AI systems. Major metaphors for input data alone, like DATA ARE RESOURCES TO BE CONSUMED and DATA ARE FORCES OF NATURE TO BE TAMED, also leave out this crucial human component (Puschmann & Burgess, 2014; Stark & Hoffmann, 2019). Yet, an AI system cannot be isolated from its input, nor from the human intentions that go into shaping that input, otherwise the system would not operate. Therefore, AI may better be understood, not as a detached set of logical operations performed on input data, but as something that is dependent upon (and encompasses) its inputs, and should be designed to share the interests of all those who take part in creating it.

Unfortunately, AI systems are not commonly designed this way, as those who benefit most tend not to be those providing the inputs, nor those subject to the outputs. AI may best be understood, perhaps then, as a social control system (Whitaker, 2021) which influences the states of its inputs and outputs. This conceptualization of AI is more in line with modern theories of the brain, which frame it as something that constructs control loops, exploiting consistencies in the environment and establishing reliable rules for interaction (Cisek, 1999).

However, scale is an important distinguishing factor here, as AI control loops encompass and exploit much larger arrangements of interacting parts (relative to the control system of a single brain). Energy cost and environmental damage of training large AI models aside (Bender et al., 2021), the social harms of AI can be perpetuated by control loops that have been constructed from data which reflect social biases, racism, ableism, and genderism. Predictive policing is a prime example of a large-scale exploitative AI control loop: i) a police organization aims to prevent crime, ii) they train an AI tool on historical data revealing that Black and Brown neighborhoods are where most arrests take place, iii) they then focus on making arrests in those neighborhoods and less in others, iv) data from those arrests are added back into the predictive policing model, v) Black and Brown neighborhoods become even more likely targets the next time the model is instantiated. An important confound with this data-driven policy, however, is that arrest data is not crime data (O'Neil, 2017). What is defined as criminal activity is socially constructed, a matter of power and subject to change, and arrests are decisions made by those in power -- they are not indications that true wrongdoing happens or not. This is common knowledge, especially to those who are victims to police violence. Yet, it is common knowledge that would be difficult (if not impossible) to encode into predictive policing even if those in power were to prioritize fair consideration of related data, which they have been historically reluctant to do.

Predictive policing with AI can thus create a control loop in which available data kickstarts a racist, data-driven social policy, which becomes further cemented with every instantiation of the AI model. This is so much so, that Ruha Benjamin suggests predictive policing "should more accurately be called crime production algorithms" (Benjamin, 2019). Rather than reducing violence in marginalized communities, hyper-surveillance paired with such algorithms can bring violence directly to those communities (Stroud, 2021). This pattern, in which AI-based prediction influences outcome, marks an imposition of determinability and control on systems which, by nature, are quite indeterminable (Birhane, 2021). Unfortunately, this feature is common to the AI control systems designed to gain influence over social environments: platform and freelance workers feel the experience of an "invisible cage", in which factors that determine their success are unknowable and unpredictable (Rahman, 2021); online artists and creators seek advice from "algorithmic experts" in hopes of gaining "algorithmic recognition" (Bishop, 2020), and alter marketing strategies according to an algorithmic "visibility game" (Arriagada & Ibáñez, 2020); students taking online exams aim to decode test-grading algorithms rather than learning the intended material (Chin, 2020); Uyghurs under constant surveillance of emotion-detection AI feel they

"have nowhere to hide" (Wakefield, 2021). To these people, AI is viewed as a controller of their landscape, as a dictator of their path. Thus not only does AI exploit consistencies in its environment, it inflicts them by forcing tech-defined norms upon the people within it. While people themselves (and their brains) are capable of large-scale social control, AI-labeled technologies make that process more efficient and readily accepted due to their perceived intelligence and objectivity. As control systems then, the brain and AI may have some similarities, but the scale and ease of influence (and potential harm) is perhaps the most irreconcilable difference.

Given the social significance of these points, we believe there are more appropriate terms for AI-labeled technologies that do not entail the Computational Metaphor. We discuss some of the initiatives to change the AI lexicon in the next section, and we hope that neuroscientists and brain researchers can look to those sources, evaluate the Computational Metaphor in new terms, and support those initiatives. We think this is important work for researchers of the brain because, as such, they have strong influence over the language that is used to describe AI-labeled technology.

## The need for a new lexicon

Pratyusha Kalluri posed an important question about AI in a recent Nature article: how does it shift power (Kalluri, 2020)? We believe that the same question deserves more attention in scientific debates about the Computational Metaphor: who gains power under the language which anthropomorphizes machine intelligence, and who loses it? With this in mind, many researchers are advocating for more appropriate AI terminology and communication strategies, and we list some of that work going on below.

The AI Now Institute, a center that studies the social implications of AI, is currently putting together a lexicon which questions critical AI concepts like fairness and transparency through a historical and social lens (Raval & Kak, 2021). One of its founders, Kate Crawford, recently put forth the metaphor AI IS AN ATLAS, which not only captures AI's purpose of finding insight through collective knowledge, it also highlights AI's colonizing purpose to become "the dominant way of seeing" and "determine how the world is measured" (Crawford, 2021). Vladan Joler and Matteo Pasquinelli, writers of Nooscope Manifesto, extend on the atlas metaphor and ultimately promote the idea that AI IS A VIEWING TOOL (Pasquinelli & Joler, 2020). In this metaphor, training data corresponds to the light beam reflecting off the thing to be observed, the learning algorithm corresponds to the lens, and the final representation corresponds to the statistical model. Conceptualizing AI as a microscope or telescope captures its purpose of finding patterns in data which are generally difficult to see, but it also gives room for AI's imperfections, just as a lens can distort or diffract light. Thomas Mullaney, Benjamin Peters, Mar Hicks, and Kavita Philip de-mystify tech-based metaphors head on (Mullaney et al., 2021). Selecting powerful and highly-marketed terms and concepts, they and other social scientists deconstruct tech metaphors to reveal hidden truths in chapters like THE CLOUD IS A FACTORY, GENDER IS A CORPORATE TOOL, and CODING IS NOT EMPOWERMENT. Interestingly, in one chapter of the book, Sarah Roberts flips the meaning of AI IS A HUMAN to show how builders of AI systems abuse human labor to make their products seem smarter than they actually are. And Luke Stark suggests FACIAL RECOGNITION IS THE PLUTONIUM OF AI, capturing the idea that its risks outweigh its benefits, it has socially toxic effects, and that it should be banned for almost all practical purposes (Stark, 2019).

By examining the purpose of AI-labeled technologies, some of these researchers expose features of AI which remain covered underneath that label. The Computational Metaphor, on the other hand, seems to be popularized based on perceived parallel features of brains and computers, rather than their respective purposes. Neuroscientist, Paul Cisek, suggested the Computational Metaphor is flawed for this very reason, using a metaphor with an analogous flaw, CARS ARE ENERGY CONVERTERS:

*"A car may be described as a device for converting chemical energy into kinetic energy. This description is not false, and could serve as the foundation for a scientific study of cars. Such study could lead to theories of how parts of the car contribute to its role of energy conversion … but it would fail to provide a complete picture of the car's function. In order to understand the purpose of such things as the steering wheel, one has to understand that the function of a car is to transport people. Energy conversion is merely a useful means toward that end,"* (Cisek, 1999)**.**

Indeed, brains and computers share some features, but whereas computation is the sole purpose of a computer, that of the brain is not entirely determined. Perhaps until we figure that out, there is no single metaphor to which we can all agree (Vlasits, 2017). But in the meantime, the Computational Metaphor deserves challenge from the neuroscience community, not only on the theoretical/clinical side as is custom, but on the social side, in order to more fully serve the public that funds them and their messaging. We hope the initiatives of the AI researchers mentioned above can inspire new lines of thinking about the Computational Metaphor.

## Conclusion

The reason for this essay is not to argue why the brain is not a computer -- at least not for the sake of neuroscience. Instead, it is a call to look beyond neuroscience, and to recognize how its most prevalent and controversial metaphor is used to market tools that are oversold, undervetted, and often harmful. As the AI and neuroscience fields continue to influence each other, the work of neuroscientists is perhaps more socially relevant now than ever in the past. Yet as experts in the study of intelligent systems, there has been little discussion about how the Computational Metaphor, and specifically the term artificial "intelligence", has been used to legitimize powerful and false ideologies that serve to diminish human and worker rights. We hope this manuscript serves as an invitation to that discussion.

Neuroscientists can address this issue by learning more about it. Read and support the work of the AI researchers above, and institutes like AI Now and Data & Society that communicate the need for a new AI lexicon. For those who use the Computational Metaphor as a central model in their lab, make time in journal clubs and lab meetings to discuss the social implications of AI, as the work of mentees and labmates may go on to inform future AI development. Talks and conference presentations can also be crafted in ways to discuss one's computational models in the context of AI's current social challenges. Conference organizers can lead this effort, similar to how the 2020 NeurIPS conference encouraged social relevance statements to be included in abstract submissions. Computational neuroscience departments can also make room for AI ethics coursework, as many computer science departments across the country now require. All of these efforts would encourage one to evaluate the language they use to describe brains, technology, and society, and inspire the usage of perhaps more enlightening metaphors and concepts.

As a closing note, author ATB was recently listening to a podcast on the scientific challenges across neuroscience and AI research (Middlebrooks, 2021). While multiple scientists cited "intelligence" as a problematic term, Jonathan Brennan perhaps best illustrated the issue:

*"Consider an AI system designed to predict the next action in a baseball game… We have tons of rich, rich data about baseball games… So feed that data to the biggest architecture that you want to handle that dynamic, spatiotemporal kind of data… Let's say it does quite well at predicting what's going to happen next in the baseball game. Can you then go into that system and extract some systematic principles that help you understand and explain how a game like baseball works? I suspect not… It's hard to imagine I can go into that neural network that has learned to predict the next baseball move and find the law of gravity. That law is not present in a way that can be understood in the neural network that has been trained to predict what's happening in the baseball game… The way we are approaching the merger of AI and neuroscience is not grounded in a goal of understanding, it's grounded in a goal of prediction: can we predict what is going to happen next? But that goal is not the same as understanding."*

In other words, while AI may offer better predictions and open new avenues to explore with more data (Hasson et al., 2020), that does not automatically translate to more insight: recording all of the neurons in the brain may not reveal a general theory of brain function; more text and speech data likely will not reveal the underlying context of words in the moment; more face data from arrested people will not explicitly reveal America's history of systemic racism. Nonetheless, the "intelligence" label of AI systems lingers as data sources and network architectures grow in size, showcasing ever more impressive performance scores while offering only an illusion of understanding. But what meaning is "intelligence" to society if understanding a problem, or having the vision to address or overcome the problem, is not part of it?

We, the authors, are also neuroscientists, and so we recognize the conceptual and logistical challenges that brain researchers may face when considering the points we have made. We are still learning how to adjust to them, ourselves. Certainly, the Computational Metaphor has inspired useful tools to the benefit of many, and in some cases it may still serve as the best metaphor for one's work. But its overuse in the tech industry is raising fields of red flags, misguiding technology development under the guise of "intelligence", and further de-centering marginalized folk who are already de-centered by tech-defined norms. While the Computational Metaphor may serve the institute of science well in some respects, it is important to consider to what degree it serves everyone else.

## Acknowledgements

The authors would like to thank Sara Berger, Nick Montfort, Luke Stark, Anna Wexler, Buddhika Bellana, and Brian Maniscalco for their comments on earlier versions of this manuscript. ATB would also like to thank Lupe Fiasco, Nikki Jean, and Full Circle at the Society of Spoken Art, without whose teachings on the power of metaphor this manuscript would not have been written.